\documentclass[showpacs,amssymb,twocolumn,aps]{revtex4}
\usepackage{amssymb}
\usepackage{amsmath}
\usepackage[pdftex]{graphicx}
\usepackage{epstopdf}
\usepackage[utf8]{inputenc}

\usepackage{fancyhdr}
\usepackage{slashed}

\begin{document}

\date{\today}
\title{QED plasma in a background of static gravitational fields}
\author{F. T. Brandt, J. Frenkel and J. B. Siqueira }
\affiliation{Instituto de F\'{\i}sica, Universidade de S\~ao Paulo,
S\~ao Paulo, SP 05315-970, Brazil}


\begin{abstract}
We derive, in $d$-dimensional space-time,
the effective Lagrangian of  static gravitational fields interacting with a QED plasma at high temperature.
Using the equivalence between the static hard thermal loops and those with zero external energy-momentum, 
we compute the effective Lagrangian up to two-loop order. We also obtain a non-perturbative contribution which arises from the sum
of all infrared divergent ring-diagrams. 
From the gauge and Weyl symmetries of the theory, we deduce to all orders that this effective Lagrangian is equivalent to the 
pressure of a QED plasma in Minkowski space-time, with the global temperature replaced by the Tolman local temperature.
\end{abstract}

\maketitle

\section{Introduction}

The search for a consistent thermal field theory in the perturbative regime has led to realization that all the 
so-called {\it hard thermal loops} have to be taken into account.
In momentum space, these are amplitudes with loop momenta of the order of the temperature, which is large compared with all the external momenta.
In  the case of gauge field theories, it has been shown that it is possible to construct
a closed form expression for the  effective Lagrangian, which generates all hard thermal 
loops \cite{Frenkel:1989br,Taylor:1990ia,Braaten:1990az}.

In gauge theories the hard thermal loop amplitudes are related to each other through Ward identities.  
This property, together with the characteristic non-localities exhibited by the amplitudes, are key ingredients for the construction of 
a gauge invariant effective action.
In principle, the same approach can be employed for hard thermal loops in a background of soft gravitational fields.
It is known that, similarly to the gauge field amplitudes, 
the graviton thermal amplitudes satisfy, in the high temperature limit, simple Ward identities
which reflect the symmetry under local coordinate transformations (in addition, these amplitudes are
also related by Weyl identities which arise from the scale invariance) \cite{Brandt:1993dk}. 
Nevertheless, an explicit closed form expression for the one-loop order effective Lagrangian is only known in two special limits,
when the background gravitational field is either time independent or spatial independent.
Each of these limits (which physically corresponds to static or long wavelength plasma perturbations)
yield two different local effective Lagrangians which are functionals 
of the background field \cite{Brandt:2009ht,Francisco:2013vg}. 
In the case of a general configuration of the background gravitational fields, 
so far only an implicit representation of the one-loop effective Lagrangian is known \cite{Brandt:1994mv}.

In previous works it has been shown that the leading contributions of hard thermal loops in a static background can be obtained
by evaluating them at zero external energy-momentum. This has been shown both at one-loop \cite{Frenkel:2009pi} 
and subsequently at two-loops \cite{Brandt:2012mn}. More recently, this result has also been generalized to all orders 
\cite{bfs2013}. 
This interesting property has prompted us to consider a more direct approach in seeking for the 
effective Lagrangian, by making use of the background field method \cite{peskin_scroeder}
(for a pedagogical review article see also \cite{abbott82}). 

As an example of the usefulness of this method, the one-loop effective Lagrangian has been previously derived in a simple manner 
in the case of thermal scalar fields in a static background \cite{Brandt:2012ei}. 
The result, which has been known for many years \cite{Rebhan:1991yr}, has the same metric dependence as in Eq. \eqref{eq11}, 
differing only by a factor which counts the degrees of freedom of the fields. 
This form exhibits some interesting properties.
First, it has a characteristic dependence on the local temperature
\begin{equation}\label{Tloc}
T_{\rm loc}\equiv \frac{T}{\sqrt{g_{00}}},
\end{equation}
being proportional to $T^d_{\rm loc}$, where $d$ is the space-time dimension and $T$ is the asymptotic Minkowski space temperature.
$T_{\rm loc}$ is the temperature measured by a standard local thermometer, such as a Carnot cycle \cite{Balazs2:1965,Ebert:1973}.  
This behavior is in agreement with the so-called {\it Tolman-Ehrenfest effect} 
which argues that in a system at thermal equilibrium in a stationary gravitational
field, the temperature varies with the space-time metric according to
the relation \eqref{Tloc}.  This effect was originally discovered in the context of
a classical fluid interacting with external static gravitational fields \cite{Tolman:1930zza,Tolman:1930a,Tolman:1930zza1}.

Another important property of the one-loop static effective Lagrangian is its invariance under conformal transformations, for any value of $d$.
This can be simply understood since, in order to behave like a density, the factor $T^d_{\rm loc}$ in the effective Lagrangian
has to be multiplied by $\sqrt{|g|}$. Consequently, the resulting expression is invariant under 
the rescaling $g_{\mu\nu} \rightarrow \sigma g_{\mu\nu}$.

The main purpose of the present work is to obtain the higher loop corrections to the effective Lagrangian of a QED plasma in a static gravitational
background. The one-loop results above described, do not take into account the interactions between electrons and photons.
In order to consider these effects, 
we will apply the same basic idea of the background field method to higher loop orders.
As we will show, in a $d$-dimensional space-time, the interactions break the Weyl symmetry when $d\neq 4$.
The two-loop contribution, given by Eq. \eqref{eq21}, is obtained computing the 1PI diagrams, with no external legs, in the background of
static gravitational fields. We also compute a non-perturbative contribution to the effective Lagrangian which arises from the summation of all higher order infrared divergent 1PI diagrams. In this case, there are two different forms, given by Eqs. \eqref{Leven} and \eqref{Lodd}, depending 
whether the space-time dimension is even or odd, respectively. 
The above results have a very simple structure. They are equivalent, up to a factor
of $\sqrt{|g|}$, to the pressure of a QED plasma at high temperature
in $d$-dimensional Minkowski space-time, with $T$ replaced by the local temperature $T_{\rm loc}$.
This shows that the Tolman-Ehrenfest effect is explicitly manifested even when
the quantum corrections are taken into account.
These and other related aspects of the effective Lagrangian are discussed further in the concluding section.

\section{One-loop effective Lagrangian}
 
In this section we will introduce our basic notation and method. Also,
for completeness we derive the one-loop effective Lagrangian.

Let us consider the Lagrangian for photons and electrons in a gravitational background
\begin{align}\label{eq:lll}
\mathcal{L} &= \sqrt{|g|} \left[ i \bar{\psi} g^{\mu \nu} \gamma_\mu(\partial_\nu -i e A_\nu) \psi  
-\frac{1}{4} g^{\mu \nu}g^{\alpha \beta} F_{\mu \alpha}F_{\nu \beta} 
\right. \nonumber \\ &
- \left. \frac{1}{2} (g^{\mu \nu} \partial_\mu A_\nu )^2 + g^{\mu \nu} \partial_\mu \bar{C} \partial_\nu C \right],
\end{align}
where $g_{\mu\nu}$ is the metric tensor ($|g| = |\det{g_{\mu\nu}}|$), 
$F_{\mu\nu} = \partial_{\mu} A_\nu - \partial_{\nu} A_\mu$ 
is electromagnetic field tensor, $\psi$ is the fermion field and $C$ is the ghost field
(we are employing the Feynman gauge condition for the gauge field $A_\mu$). As we have pointed out in the introduction, 
for the purpose of obtaining the static effective Lagrangian in the high temperature limit, 
we will neglect all the space-time derivatives of the metric as well as the fermion masses
since these quantities would be suppressed by the much larger scale of temperature. 

The Feynman rules for photons, fermions and ghosts in a gravitational background can be readily obtained
using the vierbein formalism, which allows us to write the  Lagrangian in the following form
\begin{align}\label{eq:ll123}
\mathcal{L} &= \sqrt{|g|} \left[ i \bar{\psi}  {\tilde{\gamma}}^a(\tilde{\partial}_a -i e \tilde{A}_a) \psi -\frac{1}{4}  
\tilde{F}_{a b}\tilde{F}^{ab} \right. 
\nonumber \\&-\left.\frac{1}{2} ( \tilde{\partial}_a \tilde{A}^a )^2 + \tilde{\partial}_a \bar{C} \tilde{\partial}^a C \right],
\end{align}
where $A_a = E^{\mu}_a A_\mu$ is the field in the local frame ($E^\mu_a$ is the vierbein) and the Dirac matrices $\tilde\gamma_a$ satisfy
\begin{align}\label{eq::algebra}
\{ \tilde{\gamma}_a, \tilde{\gamma}_b\} = 2 {E_a}^\mu {E_b}^\nu g_{\mu \nu} =   2 {E_a}^\mu {E_b}^\nu {e^c}_\mu {e^d}_\nu \eta_{c d} = 2\eta_{ab},
\end{align}
From this Lagrangian one obtains the following Feynman rules for the effective propagators and vertices
\begin{subequations}\label{frules}
\begin{eqnarray}\label{fprop}
\includegraphics[scale=0.8]{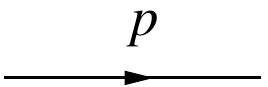} & : \;\;\; & \;\;\;
\displaystyle{{1\over \sqrt{|g|}\gamma^a p_a}} 
\end{eqnarray}
\begin{eqnarray}
\includegraphics[scale=0.8]{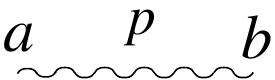}
& : \;\;\;&\;\;\; \displaystyle{{\eta_{ab}\over \sqrt{|g|} p^c p_c}} 
\end{eqnarray}
\begin{eqnarray}
\includegraphics[scale=0.8]{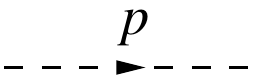}
 & : \;\;\;&\;\;\; 
-\displaystyle{{1\over \sqrt{|g|} p^a p_a}} 
\end{eqnarray}
\begin{eqnarray}
\begin{array}{c}\includegraphics[scale=0.8]{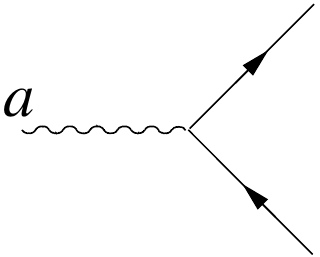}\end{array}
 & : \;\;\;&\;\;\; 
e\sqrt{|g|} \gamma^a
\end{eqnarray}
\end{subequations}

\begin{figure}[t!]
\includegraphics[scale=0.8]{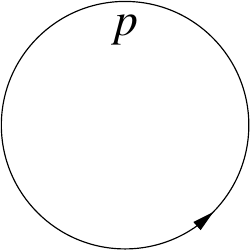}\qquad\includegraphics[scale=0.8]{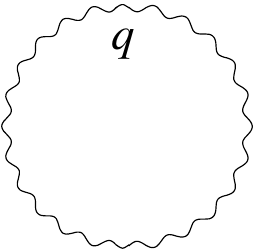}\qquad\includegraphics[scale=0.8]{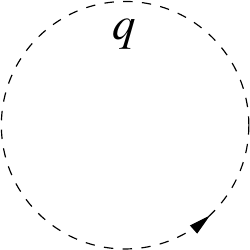}
\caption{One-loop diagrams which contribute to the effective Lagrangian.
The solid line represents a fermion and
wavy and dashed lines denote respectively gauge and ghost particles, all in the gravitational background.} \label{fig1}
\end{figure}

The lowest order contributions to the effective Lagrangian are represented diagrammatically in Fig.~\ref{fig1}.  
Let us first consider the fermion loop contribution. Using the imaginary time formalism and performing the Dirac algebra in the
vierbein basis, we obtain
\begin{align}\label{LF}
\mathcal{L}_1^F &=T \sum_{n} \int \frac{d^{d-1} p}{(2 \pi)^{d-1}} \mathrm{tr} \log ( \beta \tilde{\gamma}^a \tilde{p}_a)\nonumber\\
 &= \frac{T}{2} 2^{E(d/2)} \sum_{n}\int \frac{d^{d-1} p}{(2 \pi)^{d-1}}  \log (-\beta^2 g^{\mu \nu} {p}_\mu {p}_\nu ), 
\end{align}
were $\beta=1/T$ and we are considering a $d$-dimensional space-time. The time component of the momentum is 
$p_0 = i \omega^F_n$, where $\omega^F_n =  (2n+1)\pi T$ are the fermionic Matsubara frequencies and $n=0,\pm 1, \pm 2, \dots$.
Here we are employing  the minimal representation for the Dirac matrices, so that the trace of the identity is given by
$\mathrm{tr} I =2^{E(d/2)}$, where $E(d/2)$ is the integer part of $d/2$.

Similarly the respective contributions from the photon and ghost loops shown in Fig.~\ref{fig1} can be expressed as follows
\begin{subequations}\label{LB}
\begin{eqnarray}
\mathcal{L}_1^P &=& -\frac{d}{2} T \sum_n \int\frac{d^{d-1} q}{(2\pi)^{d-1}} \log( -\beta^2 g^{\mu \nu} q_\mu q_\nu) \\
\mbox{and} &&\nonumber \\
\mathcal{L}_1^G &=& T \sum_n \int\frac{d^{d-1} q}{(2\pi)^{d-1}} \log( -\beta^2 g^{\mu \nu} q_\mu q_\nu),
\end{eqnarray}
\end{subequations}
where $q_0 = i \omega^B_n$ and $\omega^B_n =  2 n \pi T$ is the bosonic Matsubara frequency , with
$n=0,\pm 1, \pm 2, \dots$.

In order to perform the sum/integrals in Eqs. \eqref{LF} and \eqref{LB} we now use a locally rest     
vierbein frame as defined in the appendix \ref{aa}. 
Proceeding in this way, Eq. \eqref{LF} yields
\begin{align}
\mathcal{L}_1^F &= \frac{T}{2}2^{E(d/2)} \sum_{n} \int \frac{d^{d-1} p}{(2 \pi)^{d-1}}  \log \left\{ -\beta^2 \left[  (p_0/\sqrt{g_{00}})^2 \right. \right. 
\nonumber \\ &+ \left. \left. g^{ij}(p_i+ (g^{-1})_{ik}g^{0k} p_0 ) (p_j+ (g^{-1})_{jl} g^{0l} p_0) \right] \right\}.
\end{align}
Making a change of variables in the $p_i$ integration, it is possible the factorize all the metric dependence and we obtain the following
result
\begin{align}
\mathcal{L}_1^F &=  2^{E(d/2)} \frac{ \sqrt{(-1)^{d-1} \bold{g}^{-1} g_{00} }}{g_{00}^{d/2}}  \,\, \frac{T}{2} \sum_{n} \int \frac{d^{d-1} p}{(2 \pi)^{d-1}} 
\nonumber \\ &\times \log \left\{ -\beta^2 \left[  p_0^2 - |\vec{p}|^2 \right] \right\}.
\label{eq:P_0f234aaa}
\end{align}
where $\bold{g}^{-1} = \det (g^{-1})_{ij}$. The sum/integral in the previous expression is the same as in 
flat space-time \cite{kapusta:book89}, here generalized to $d$ space-time dimensions. 
The temperature-independent part of \eqref{eq:P_0f234aaa} leads to a divergent integral.
This divergent result gives just the zero-point energy of the vacuum,
which can be subtracted off since it is an  unobservable constant.
On the other hand, the $T$-dependent part of \eqref{eq:P_0f234aaa} leads to a finite result.

The metric dependence 
can be dealt with expanding the determinant of the metric in terms of co-factors (see appendix \ref{apb}). 
In this way, the fermionic contributions to the one-loop effective Lagrangian reduces to the following expression
\begin{equation}
\mathcal{L}_1^F =  \sqrt{|g|} \left(\frac{T}{\sqrt{g_{00}}}\right)^d \frac{ \Gamma(d) \zeta(d) 2^{E(d/2)}(1-2^{1-d}) }{(2\sqrt{\pi})^{d-1} \Gamma(\frac{d+1}{2})} ,
\label{eq:p0f}
\end{equation}
where $\Gamma$ and $\zeta$ are the Euler and Riemann functions, respectively.

Proceeding similarly, the sum of the photon and ghost contributions in \eqref{LB} yields
\begin{equation}\label{eq:p0b}
\mathcal{L}_1^B = \sqrt{|g|} \left(\frac{T}{\sqrt{g_{00}}}\right)^d (d-2)  \frac{\Gamma(d) \zeta(d)}{(2 \sqrt{\pi})^{d-1} \Gamma(\frac{d+1}{2})}.
\end{equation}
Finally, adding together the results \eqref{eq:p0f}  and \eqref{eq:p0b}, we obtain the one-loop effective Lagrangian in the form
\begin{align}\label{eq11}
\mathcal{L}_1 &= \sqrt{|g|} \left(\frac{T}{\sqrt{g_{00}}}\right)^d  \frac{\Gamma(d) \zeta(d)}{(2 \sqrt{\pi})^{d-1} \Gamma\left(\frac{d+1}{2}\right)} \nonumber \\ &\times \left[(d-2)+2^{E(d/2)}(1-2^{1-d}) \right].
\end{align}
As expected for a density, Eq. \eqref{eq11} exhibits the factor $\sqrt{|g|}$.  It also displays a temperature dependence in terms of the 
{\it local temperature} defined in Eq. \eqref{Tloc}, which is a simple consequence of the Tolman-Ehrenfest effect.
The combination of these two factors leads to the invariance under the scale transformation $g_{\mu\nu} \rightarrow \sigma g_{\mu\nu}$.
In the following sections we will investigate the effect of higher order corrections, when the thermal photons interact also with  thermal fermions.

\section{Effective Lagrangian at two-loop order}

\begin{figure}
\center
  \includegraphics[scale=1.0]{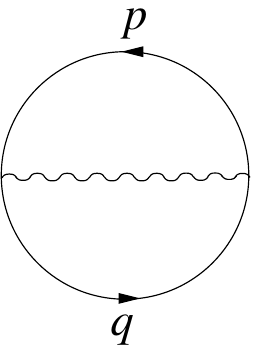}\\
  \caption{Two-loop contribution to the effective Lagrangian. The effective vertex and propagators are given by
Eqs. \eqref{frules}}\label{2-loops}
\end{figure}

Let us now apply the technique illustrated in the simple one-loop calculation of the previous section, in order to obtain
the two-loop order contribution to the effective Lagrangian. This can be obtained computing the diagram 
shown in Fig.~\ref{2-loops}. Using the effective Feynman rules given in \eqref{frules} we obtain
\begin{align}
\mathcal{L}_2 &= \frac{e^2T^2}{2} \sum_{m,\, n} \int \frac{d^{d-1}p}{(2\pi)^{d-1}} \frac{d^{d-1}q}{(2\pi)^{d-1}} \nonumber \\ & \times \mathrm{tr} \left( {\gamma}^\mu \frac{1}{{\gamma}^\nu p_\nu} {\gamma}_\mu \frac{1}{{\gamma}^\alpha p_\alpha} \right) \frac{1}{\sqrt{|g|}(p - q)^\gamma(p - q)_\gamma},
\end{align}
where $q_0 = i  (2n+1)\pi T$  and $p_0 = i (2m+1) \pi T$. In the vierbein basis the Dirac algebra yields
\begin{align}
\mathcal{L}_2 &= \frac{e^2T^2}{2\sqrt{|g|}} 2^{E(d/2)}(2-d) \sum_{m,\, n} \int \frac{d^{d-1}p}{(2\pi)^{d-1}} \frac{d^{d-1}q}{(2\pi)^{d-1}}  \nonumber \\  &\times \frac{\tilde{p}^a \tilde{q}_a }{(\tilde{p}^b \tilde{p}_b) (\tilde{q}^c \tilde{q}_c)(\tilde{p}-\tilde{q})^d(\tilde{p}-\tilde{q})_d },
\end{align}
which can be rewritten as
\begin{align}
\mathcal{L}_2 
&=\frac{e^2 2^{E(d/2)}(2-d)}{4\sqrt{|g|}} 
\nonumber \\ &\times  \left[2T^2 \sum_{m,\,  l} \int \frac{d^{d-1}p}{(2\pi)^{d-1}} \frac{d^{d-1}k}{(2\pi)^{d-1}}  \frac{1}{(\tilde{p}^a \tilde{p}_a) (\tilde{k}^b \tilde{k}_b)}
\right. \nonumber \\ & \left.
- T^2\sum_{m,\, n} \int \frac{d^{d-1}p}{(2\pi)^{d-1}} \frac{d^{d-1}q}{(2\pi)^{d-1}}  \frac{1}{(\tilde{p}^a\tilde{p}_a)(\tilde{q}^b \tilde{q}_b)} \right]  ,
\label{eq:p2fac}
\end{align}
where $k=p-q$ is the photon momentum ($k_0= i 2 l \pi$). There are two independent sum/integrals in Eq. \eqref{eq:p2fac}, namely
\begin{equation}
I_1 = T \sum_{m} \int \frac{d^{d-1}p}{(2 \pi)^{d-1}} \frac{1}{\tilde{p}_a \tilde{p}^a }
\end{equation}
and
\begin{equation}
I_2 = T \sum_{l} \int \frac{d^{d-1}k}{(2 \pi)^{d-1}} \frac{1}{\tilde{k}_b \tilde{k}^b},
\end{equation}
which differ since $k_0= i 2\pi l T$ and $p_0 = i \pi (2m+1) T$.

Using the definition of the locally rest vierbein \eqref{eq:vierbein-final}, the integral $I_1$ can be written as
\begin{widetext}
\begin{align}
I_1 &=T \sum_{m} \int \frac{d^{d-1}p}{(2 \pi)^{d-1}} \frac{1}{ (p_0/\sqrt{g_{00}})^2 + g^{ij}[p_i - (g^{-1})_{ik} g^{0k} p_0][p_j - (g^{-1})_{jl} g^{0l} p_0]}.
\end{align}
\end{widetext}
Proceeding as in the one-loop case, a change of variables in $p_i$  allows one to perform  the thermal part of 
the resulting integral, which yields the result 
\begin{align}
I_1 &= \frac{\sqrt{|g|}}{(g_{00})^{d/2-1}} T^{d-2}(2-2^{4-d}) \frac{\Gamma(d-2) \zeta(d-2)}{\Gamma(\frac{d-1}{2})(2\sqrt{\pi})^{d-1}}.
\label{eq:i1-final}
\end{align}
Similarly, we obtain the following result for the integral $I_2$
\begin{align}
I_2 &= -2 \frac{\sqrt{|g|}}{(g_{00})^{d/2-1}} T^{d-2} \frac{\Gamma(d-2) \zeta(d-2)}{\Gamma(\frac{d-1}{2})(2\sqrt{\pi})^{d-1}} .
\label{eq:i2-final}
\end{align}
Substituting \eqref{eq:i1-final} and \eqref{eq:i2-final} in Eq. \eqref{eq:p2fac}, yields
\begin{align}\label{eq21}
\mathcal{L}_2 &= \sqrt{|g|} \left(\frac{T}{\sqrt{g_{00}}}\right)^d \left[e^2 \left(\frac{T}{\sqrt{g_{00}}}\right)^{d-4} \right] 2^{E(d/2)-2}(2-d) \nonumber \\ &\times   \left[\frac{\Gamma(d-2) \zeta(d-2)}{\Gamma(\frac{d-1}{2})(2\sqrt{\pi})^{d-1}}\right]^2  (2-2^{4-d})(6-2^{4-d}).
\end{align}
This result, which  can be identified with the two-loop contribution to the pressure, 
in the high temperature limit, is similar to the flat space-time result,
corrected by the factor $\sqrt{|g|}$ (as expected for a density).  Also,
the two-loop result exhibit the simple dependence on the local temperature, as defined in \eqref{Tloc}.

\section{Non-perturbative contribution to the effective Lagrangian}
\begin{figure}
  \includegraphics[scale=0.5]{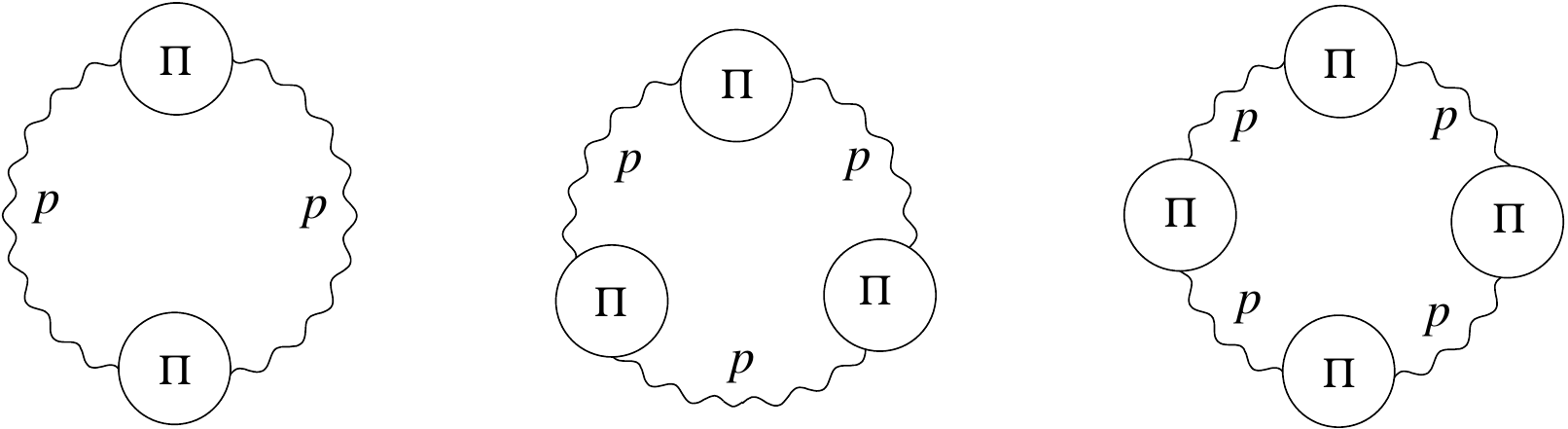}
  \caption{Ring diagrams.}\label{rings}
\end{figure}

Higher loop corrections to the effective Lagrangian may exhibit infrared divergences which arise from the dominant high temperature contribution of the zero mode.
In order to deal with these divergences one has to sum an infinite series of diagrams which are individually divergent.  
Each successive order is obtained by inserting an extra one-loop static photon self-energy 
(the photon self-energy diagram is shown in Fig. \ref{pi-foton_g} of the appendix \ref{apc}). 
By power counting one can see that these zero mode contributions will produce infrared divergences in the high temperature limit, 
when the number of insertions is large enough.
For instance, in four space-time dimensions diagrams with two or more self-energy insertions are infrared divergent.

Figure \ref{rings} shows three such diagrams, corresponding to three-, four- and five-loops.
Diagrams in this set are called {\it ring diagrams}.
A diagram with $N$ self-energy insertions has the form 
\begin{widetext}
\begin{align}
R_N &=  \frac{T}{2N} \int \frac{d^{d-1}p}{(2\pi)^{d-1}} \left[ (-1)^N\tilde{D}^{a_1 b_1} \tilde{\Pi}_{b_1 a_2} \cdots \tilde{\Pi}_{b_{N-1} a_{N}} \tilde{D}^{a_N b_N} \tilde{\Pi}_{b_N a_1} \right]_{\tilde{p}_0=0} ,
\end{align}
where $D^{a b}$ is the photon propagator. Using the Eq. \eqref{eq:static-self-energy} for the static self-energy we obtain
\begin{align}
R_N&=  \frac{T}{2N} \int \frac{d^{d-1}p}{(2\pi)^{d-1}} \left[ \frac{2m^2}{\sqrt{|g|} g^{ij} p_i p_j}\right]^N ,
\label{eq:jebun}
\end{align}
where $m$ is the thermal mass given by \eqref{Tmass}.
Performing a  change of variables, we obtain
\begin{align}
R_N &= \left( \frac{2 m^2}{\sqrt{|g|} }\right)^{(d-1)/2} T \sqrt{(-1)^{d-1} \bold{g}^{-1} }  \int \frac{d^{d-1} p}{(2 \pi)^{d-1}} \frac{1}{2N}\left[ \frac{1}{\eta^{ij} p_i p_j}\right]^N \nonumber \\
&= \sqrt{|g|} \left( \frac{2 m^2}{\sqrt{|g|} }\right)^{(d-1)/2} \frac{T}{\sqrt{g_{00}}}   \int \frac{d^{d-1} p}{(2 \pi)^{d-1}} \frac{1}{2N}\left[ \frac{1}{\eta^{ij} p_i p_j}\right]^N,
\label{eq:rnrn}
\end{align}
\end{widetext}
where we have used Eq. \eqref{detg} (from now on $p_i$ are dimensionless variables).
Simple power counting shows that these individual ring diagrams contributions are infrared divergent for  $N\geq d/2$ or $N \geq (d-1)/2$ respectively when $d$ is even or odd
(notice that the number of loops is $N+1$).  Because of this different behavior, one has to consider separately the two cases.

The sum of all  the infrared divergent contributions can be written as
\begin{widetext}
\begin{align}\label{eq25aa}
\mathcal{L}_{ring} &= \sum_{N=E(d/2)}^\infty R_N = \sqrt{|g|} \left( \frac{2 m^2}{\sqrt{|g|} }\right)^{(d-1)/2} \frac{T}{\sqrt{g_{00}}} \frac{1}{2^{d-2} \pi^{(d-1)/2} \Gamma \left( \frac{d-1}{2} \right) }  \sum_{N=E(d/2)}^\infty \int_0^\infty dp \frac{p^{d-2}}{2N} \left[ \frac{-1}{p^2} \right]^N.
\end{align}
Using integration by parts, we can write 
\begin{eqnarray}
 \sum_{N=E(d/2)}^\infty \int_0^\infty dp \frac{p^{d-2}}{2N} \left[ \frac{-1}{p^2} \right]^N =
\sum_{N=E(d/2)}^\infty \left[ \left. \frac{p^{d-1}}{2N(d-1)} \frac{(-1)^N}{p^{2N}} \right|_0^\infty + \int_0^\infty dp \frac{p^{d-1}}{d-1} \frac{(-1)^N}{p^{2N+1}} \right] .
 \label{eq:intgeral}
\end{eqnarray}
One can readily verify that the surface term vanishes when $d$ is an even number. First, for $p\rightarrow \infty$  it is immediate that
\begin{equation}
\sum_{N=d/2}^\infty  \frac{p^{d-1}}{2N(d-1)} \frac{(-1)^N}{p^{2N}} 
\longrightarrow 0 .
\end{equation}
When $p\rightarrow 0$ we obtain
\begin{align}
\sum_{N=d/2}^\infty  \frac{p^{d-1}}{2N(d-1)} \frac{(-1)^N}{p^{2N}} = \frac{p^{d-1}}{2(d-1)}\log(1+1/p^2) +  \sum_{N=1}^{d/2-1}  \frac{p^{d-1}}{2N(d-1)} \frac{(-1)^N}{p^{2N}} \longrightarrow 0. 
\end{align}
Therefore, Eq. \eqref{eq:intgeral}  with even values of $d$ reduces to 
\begin{align}
 \sum_{N=d/2}^\infty \int_0^\infty dp \frac{p^{d-2}}{2N} \left[ \frac{-1}{p^2} \right]^N = \frac{(-1)^{d/2}}{d-1} \int_0^\infty \frac{dp}{p^2} \frac{1}{1+1/p^2} =  \frac{\pi (-1)^{d/2}}{2(d-1)}.
 \label{eq:sum-ring-even}  
\end{align}
Substituting the Eq. \eqref{eq:sum-ring-even} into  Eq. \eqref{eq25aa}, we obtain the following non-perturbative contribution for even space-time dimensions
\begin{align}\label{Leven}
\mathcal{L}_{ring}^{d-even} &= \sqrt{|g|} \frac{T}{\sqrt{g_{00}}} \frac{(-1)^{d/2}}{2^d \pi^{(d-3)/2} \Gamma\left( \frac{d+1}{2} \right) } \left[ \frac{2 m^2}{\sqrt{|g|}} \right]^{(d-1)/2}  \nonumber \\
&= \sqrt{|g|} \left(\frac{T}{\sqrt{g_{00}}}\right)^d \left[e^2 \left(\frac{T}{\sqrt{g_{00}}}\right)^{d-4} \right]^{(d-1)/2}  \frac{(-1)^{d/2}}{2^d \pi^{(d-3)/2} \Gamma\left( \frac{d+1}{2} \right) } \left[ \frac{2^{E(d/2)+1}(1-2^{3-d}) \Gamma(d-1) \zeta(d-2)}{\Gamma\left(\frac{d-1}{2}\right) (2 \sqrt{\pi})^{d-1}} \right]^{(d-1)/2} ,
\end{align}
which exhibits a non-analyticity in the coupling constant of the form $(e^2)^{(d-1)/2}$. 

Let us now consider the case when $d$ is odd. In this case, one can show that the surface term in \eqref{eq:intgeral} does not vanish. Indeed, although
in the limit $p\rightarrow 0$ we obtain
\begin{align}\label{eqodd1}
\sum_{N=(d-1)/2}^\infty  \frac{p^{d-1}}{2N(d-1)} \frac{(-1)^N}{p^{2N}} = \frac{p^{d-1}}{2(d-1)}\log(1+1/p^2) +  \sum_{N=1}^{(d-3)/2}  \frac{p^{d-1}}{2N(d-1)} \frac{(-1)^N}{p^{2N}} \longrightarrow 0,
\end{align} 
when $p\rightarrow \infty$ we are left with the following finite contribution 
\begin{align}\label{eqodd2}
\sum_{N=(d-1)/2}^\infty  \frac{p^{d-1}}{2N(d-1)} \frac{(-1)^N}{p^{2N}} \longrightarrow \frac{(-1)^{(d-1)/2}}{(d-1)^2}.
\end{align}

Substituting Eqs. \eqref{eqodd1}  and \eqref{eqodd2}  into Eq. \eqref{eq:intgeral} we obtain
\begin{align}
\sum_{N=E(d/2)}^\infty \int_0^\infty dp \frac{p^{d-2}}{2N} \left[ \frac{-1}{p^2} \right]^N &= \frac{(-1)^{(d-1)/2}}{(d-1)^2} + \sum_{N=(d-1)/2}^\infty  \int_0^\infty dp \frac{p^{d-1}}{d-1} \frac{(-1)^N}{p^{2N-1}}  \nonumber \\ &= \frac{(-1)^{(d-1)/2}}{(d-1)} \left[ \frac{1}{d-1} + \int_0^\infty \frac{dp}{p} \frac{1}{1+1/p^2} \right].
\end{align}
The resulting integral is logarithmically increasing at large momenta, where
the approximations used for the ring diagrams are no longer valid. In order to
regularize this behavior, we will employ a cut-off  $\sqrt{-g^{ij} P_i P_j}$ in the
momentum of the original integral in Eq. \eqref{eq:jebun}, 
where $P$ is naturally of the same order as the local temperature.
In terms of the parameter $\mu = \sqrt{-\eta^{ij} P_i P_j}$, we then get
%
%
\begin{align}
\sum_{N=E(d/2)}^\infty \int_0^\infty dp \frac{p^{d-2}}{2N} \left[ \frac{-1}{p^2} \right]^N &= \frac{(-1)^{(d-1)/2}}{2(d-1)} \left[ \frac{2}{d-1} +  \log \left( 1+ \frac{\mu^2 \sqrt{|g|}}{2 m^2} \right) \right],
\end{align}
which yields the following expression for the non-perturbative contribution to the effective Lagrangian
\begin{align}\label{Lodd}
\mathcal{L}_{ring}^{d-odd} &=  \sqrt{|g|} \left( \frac{2 m^2}{\sqrt{|g|} }\right)^{(d-1)/2} \frac{T}{\sqrt{g_{00}}} \frac{(-1)^{(d-1)/2} }{2^{d} \pi^{(d-1)/2} \Gamma \left( \frac{d+1}{2} \right) }  \left[ \frac{2}{d-1} +  \log \left( 1+ \frac{\mu^2 \sqrt{|g|}}{2 m^2} \right) \right] \nonumber \\ 
&= \sqrt{|g|} \left(\frac{T}{\sqrt{g_{00}}}\right)^d \left[e^2 \left(\frac{T}{\sqrt{g_{00}}}\right)^{d-4} \right]^{(d-1)/2}  \frac{(-1)^{(d-1)/2} }{2^{d} \pi^{(d-1)/2} \Gamma\left( \frac{d+1}{2} \right) } \left[ \frac{2^{E(d/2)+1}(1-2^{3-d}) \Gamma(d-1) \zeta(d-2)}{\Gamma\left(\frac{d-1}{2}\right) (2 \sqrt{\pi})^{d-1}} \right]^{(d-1)/2}  \nonumber \\ & \times 
\left\{ \frac{2}{d-1} +  \log \left[ 1+ \mu^2  \left(\frac{T}{\sqrt{g_{00}}}\right)^{-2} \left(e^2 \left(\frac{T}{\sqrt{g_{00}}}\right)^{d-4} \right)^{-1}
\frac{\Gamma\left(\frac{d-1}{2}\right) (2 \sqrt{\pi})^{d-1}}{2^{E(d/2)+1}(1-2^{3-d}) \Gamma(d-1) \zeta(d-2)}
\right]   \right\}.
\end{align}
\end{widetext}
This expression has a logarithmic non-analyticity in the coupling constant $e^2$. The same type of non-analyticities have been found previously 
in the context of scalar fields in flat backgrounds \cite{Brandt:2012mu}.

As in the previous results for the one- and two-loop contributions to the effective action, 
the non-perturbative results in this section given by Eqs. \eqref{Leven} and \eqref{Lodd}  can be expressed in terms of 
the local temperature as defined in Eq. \eqref{Tloc}. This confirms that the Tolman-Ehrenfest effect is explicitly
manifested even in the extreme case when an infinite number of interactions are taken into account.

\section{Discussion}

In the present work, we have employed the equivalence between static and zero energy-momentum thermal amplitudes, 
which holds for the leading contributions at high temperature. Using this correspondence, 
we have obtained the effective Lagrangian of static gravitational fields interacting with a plasma of photons and electrons 
at high temperature, up to two-loops order. 
We have also obtained a non-perturbative contribution from the sum of the infinite set of ring diagrams.
This generalizes the previous results for the static effective action which were known only at one-loop order \cite{Brandt:2012ei}. 

It is interesting to remark that 
the contributions generated by the static gravitational fields correspond to those obtained 
for the pressure of a QED plasma
in Minkowski space-time in the following simple way:
apart from an overall factor of $\sqrt{|g|}$ which is required by gauge invariance, the only modification involves
the replacement of the Minkowski temperature by the local temperature \eqref{Tloc}. 

From a physical point of view, this universal behavior (which has also been derived 
using other approaches \cite{Balazs2:1965,Ebert:1973,Rovelli:2010mv,Haggard:2013fx}) 
can be traced back to the requirement of thermal equilibrium in a gravitational field. Indeed, the emergence of a local temperature, and consequently a 
temperature gradient, is unavoidable in thermal equilibrium to prevent heat (which interacts with gravity) to flow from regions of higher 
to those of lower gravitational potential.

Another salient feature is that the conformal invariance of the effective Lagrangian, 
which is present at one-loop order for any space-time dimension $d$,
is not in general satisfied by the higher loop corrections  (this is also the case of the one-loop 
photon self-energy given by Eq. \eqref{eq:static-self-energy}).
Both in the two-loop correction in Eq. \eqref{eq21} as well as in the non-perturbative contributions in Eqs. \eqref{Leven} and \eqref{Lodd} 
(which receives contributions of an infinite number of photon self-energy insertions), we find terms like
\begin{align}\label{eq36}
\left[ e^2 \left( \frac{T}{\sqrt{g_{00}}} \right)^{d-4} \right]
\end{align}
which will not be conformal invariant in general. 
The physical reason for this behaviour may be understood by noting that
$e^2$ is a dimensionful quantity with canonical mass dimension $4-d$.

We finally remark that the above leading results at high temperature were obtained by neglecting all masses compared with the
temperature. In four dimensions, since $e^2$ is dimensionless, these results should therefore be scale invariant. In this case, because
$\sqrt{|g|} T_{\rm loc}^4$ is invariant under scale transformations, we see that the modification $T\rightarrow T_{\rm loc}$ is the only 
possibility
which is consistent with the Weyl symmetry. When $d\neq 4$, the leading thermal results are no longer scale invariant 
(see Eq. \eqref{eq36}), but the violation of the Weyl symmetry still occurs according to the simple prescription
$T\rightarrow T_{\rm loc}$, which enforces a smooth behaviour when $d\rightarrow 4$.
Based on the above physical considerations and explicit calculations we conclude that, to all orders,
the simple correspondence $T\rightarrow T_{\rm loc}$ leads to the effective action of static gravitational fields
interacting with a QED plasma at high temperature.
Moreover, the terms involving $T_{\rm loc}$  ensure the invariance of this action under time-independent 
local coordinate transformations. 

%
Our treatment in arbitrary space-time dimensions was motivated by various unified field  theories
of gravitational, electromagnetic and other interactions, which have
been often formulated in higher dimensions.
But at present, we can indicate a direct physical application of the above
results only when $d=4$. These results may then be useful to calculate, for
example, the pressure in a plasma of electrons and photons
surrounding a hot star.

\acknowledgments
We would like to thank FAPESP and CNPq (Brazil) for a grant.
J. F. is indebted to Prof. J. C. Taylor for a helpful correspondence.

\appendix

\section{Vierbein in the presence of a thermal bath}\label{aa}

A local Lorentz frame can be defined in terms of the vierbein ${e^a}_\mu$ (also known as a tetrad), so that
in a given point of the manifold the metric can be written as \cite{BirrelDavies} 
\begin{align}
g_{\mu \nu} = {e^a}_\mu {e^b}_\nu \eta_{ab},
\end{align} 
where the Greek and Latin indices stand for general and local coordinates, respectively.
At finite temperature, the thermal bath introduces a privileged reference frame which is characterized
by its four velocity $u^\mu$.
In all points of the manifold, we have a special coordinate system, called locally rest frame, in which the four velocity of the thermal bath has the simple form
\begin{align}
u^\mu \dot{=} \left(\frac{1}{\sqrt{g_{00}}}, \vec{0} \right), 
\label{eq:u-primeira}
\end{align}
where $\dot{=}$ indicates that we are  considering the components of the vector $u$ in this particular frame.

In our notation, the components of an arbitrary vector $p$ is represented by
\begin{align}
\tilde{p}^a = {e^a}_\mu p^\mu, 
\end{align}
so that the scalar product with the thermal bath vector has the form
\begin{align}
\tilde{p}_a  \tilde{u}^a = p_\mu u^\mu \dot{=} \frac{p_0}{\sqrt{g_{00}}}.
\label{eq:p-dot-u}
\end{align}
This equation can be used to define a special class of vierbein. 
To see this, note that the Eqs. \eqref{eq:u-primeira} and \eqref{eq:p-dot-u} imply that exist a vierbein which is locally 
at rest in relation to thermal bath, in which
\begin{align}
\tilde{p}_0 \dot{=} \frac{p_0}{\sqrt{g_{00}}}.
\end{align}
Therefore, for two arbitrary vectors, we have the scalar product
\begin{align}
g^{\mu \nu} p_{\mu} q_{\nu} \dot{=} \eta^{ab} \tilde{p}_a\tilde{q}_b \dot{=} \frac{p_0 q_0}{g_{00}} + \eta^{ij} \tilde{p}_i \tilde{q}_j.
\label{eq:p-dot-q}
\end{align}
Using the identity, 
\begin{align}
\frac{1}{g_{00}} = g^{00} - (g^{-1})_{ij} g^{0i} g^{0j},
\end{align}
we can rewrite Eq. \eqref{eq:p-dot-q} in the form
\begin{align}
g^{\mu \nu} p_{\mu} q_{\nu} \dot{=} \left[g^{00} - (g^{-1})_{ij} g^{0i} g^{0j} \right] p_0 q_0   +\eta^{ij}   \tilde{p}_i \tilde{q}_j,
\end{align}
where $(g^{-1})_{ij}$ is the inverse of the spatial part of the metric
\begin{align}
(g^{-1})_{ij} g^{jl} = \delta_i^l.
\end{align}
On the other hand, we have
\begin{align}\label{aaa}
g^{\mu \nu} p_\mu q_\nu &= g^{00}p_0 q_0 + g^{0i}(p_0q_i+p_iq_0) +g^{ij}p_iq_j \nonumber \\ &=   \left[g^{00} - (g^{-1})_{ij} g^{0i} g^{0j} \right] p_0 q_0 \nonumber \\ &+ g^{ij}[p_i + (g^{-1})_{il}g^{0l}p_0][q_j +(g^{-1})_{jk}g^{0k}q_0].
\end{align}
Eqs. \eqref{eq:p-dot-q} and \eqref{aaa} allows one to define
\begin{align} \label{eq:vierbein-final}
\left\{ \begin{array}{l}
 \tilde{p}_0 = \frac{p_0}{\sqrt{g_{00}}} \\
 \tilde{p}_i = N_i^{\;k} [p_k + (g^{-1})_{kl}g^{0l}p_0], 
\end{array} \right.
\end{align}
where the symmetric matrix $N_i^{\;k}$ is  such that 
\begin{align}
g^{ij}= N^i_{\; l}\,\eta^{lm}\,N_m^{\;\; j}.
\end{align}

\section{Useful identity for the determinant}\label{apb}

For completeness, we present here a simple identity which has been deduced previously \cite{Brandt:2012ei}.  
Let us first denote the determinant of the spacial part as well as  the full determinant of the metric as follows 
\begin{subequations}
\begin{eqnarray}
\bold{g}&=& \det g^{ij} 
\\
g &=& \det g_{\mu \nu}.
\end{eqnarray}
\end{subequations}
Expanding in terms of co-factors
\begin{align}
g^{-1} = \sum_{\mu} g^{\mu 0} C_{\mu 0}
\label{eq:detapp1}
\end{align}
and using the relations
\begin{subequations}
\begin{eqnarray}
C_{00} &=& \bold{g} ,
\\
C_{0i}&=& g^{-1} g_{0i}
\end{eqnarray}
\end{subequations}
Eq. \eqref{eq:detapp1} yields 
\begin{align}
g^{-1} = g^{00}\bold{g}  +  g^{0i}g_{i0} g^{-1}.
\end{align}
Solving for $g^{-1}$ we obtain
\begin{align}
g^{-1} = \frac{g^{00} \bold{g}}{1-g^{0i}g_{0i}}.
\end{align}
Finally, using the identity
\begin{align}
g^{00}g_{00} = 1- g^{0i}g_{0i},
\end{align}
we obtain
\begin{align} \label{detg}
\bold{g}^{-1} g_{00} = g.
\end{align}

\section{Photon self-energy}\label{apc}
\begin{figure}
  \includegraphics[scale=0.6]{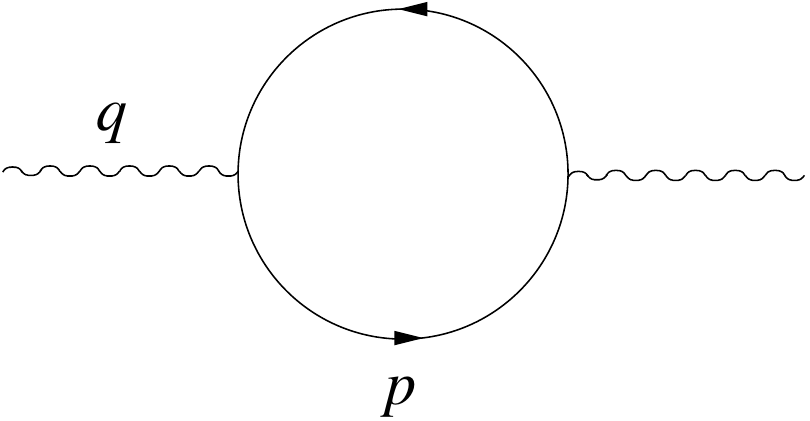}\\
  \caption{The photon self-energy}\label{pi-foton_g}
\end{figure}

The one-loop photon self-energy is shown in Fig~\ref{pi-foton_g}. The corresponding analytic expression has the form
\begin{equation}
\tilde{\Pi}^{a b}(\tilde{q}) = {e^2} T\sum_{n} \int \frac{d^{d-1}p}{(2\pi)^{d-1}} \mathrm{tr}\left(\tilde{\gamma}^a \frac{1}{\tilde{\slashed{p}}-\tilde{\slashed{q}}} \tilde{\gamma}^b \frac{1}{\tilde{\slashed{p}}}\right).
\end{equation}
Computing the trace  using the Clifford algebra \eqref{eq::algebra}, we obtain
\begin{align}
\tilde{\Pi}^{a b}(\tilde{q}) &= e^2 2^{E(d/2)}\,T\sum_{n} \int \frac{d^{d-1}p}{(2\pi)^{d-1}} \nonumber \\ &\times \frac{\tilde{p}^a(\tilde{p}-\tilde{q})^b + \tilde{p}^b (\tilde{p}-\tilde{q})^a - \tilde{p}^c (\tilde{p}-\tilde{q})_c \eta^{ab}}{\tilde{p}^2 (\tilde{p}-\tilde{q})^2} .
\label{eq:pidetr}
\end{align}
Using the definition of the locally rest vierbein \eqref{eq:vierbein-final}, we can rewrite 
\begin{widetext}
\begin{align}
\tilde{\Pi}^{ab}(\tilde{q})&=e^2 2^{E(d/2)} T\sum_{n} \int \frac{d^{d-1}p}{(2\pi)^{d-1}} \frac{\tilde{p}^a(\tilde{p}-\tilde{q})^b + \tilde{p}^b (\tilde{p}-\tilde{q})^a - \tilde{p}^c (\tilde{p}-\tilde{q})_c \eta^{ab}}{\tilde{p}^2 (\tilde{p}-\tilde{q})^2} \nonumber \\
&=e^2 2^{E(d/2)} T\sum_{n} \int \frac{d^{d-1}p}{(2\pi)^{d-1}} \frac{\tilde{p}^a(\tilde{p}-\tilde{q})^b + \tilde{p}^b (\tilde{p}-\tilde{q})^a - \tilde{p}^c (\tilde{p}-\tilde{q})_c \eta^{ab}}{\left[\frac{p_0^2}{g_{00}} - \tilde{p}_i \tilde{p}_i\right]\left[\frac{(p_0-q_0)^2}{g_{00}} -( \tilde{p}_i-\tilde{q}_i) (\tilde{p}_i-\tilde{q}_i)\right]} \nonumber \\
&= e^2 2^{E(d/2)} (g_{00})^2 T\sum_{n} \int \frac{d^{d-1}p}{(2\pi)^{d-1}} \frac{\tilde{p}^a(\tilde{p}-\tilde{q})^b + \tilde{p}^b (\tilde{p}-\tilde{q})^a - \tilde{p}^c (\tilde{p}-\tilde{q})_c \eta^{ab}}{\left[p_0^2 - g_{00} \tilde{p}_i \tilde{p}_i\right]\left[(p_0-q_0)^2 -g_{00}( \tilde{p}_i-\tilde{q}_i) (\tilde{p}_i-\tilde{q}_i)\right]} \nonumber \\
&= e^2 2^{E(d/2)} g_{00}T \sum_{n} \int \frac{d^{d-1}p}{(2\pi)^{d-1}} \frac{\bar{p}^a(\bar{p}-\bar{q})^b + \bar{p}^b (\bar{p}-\bar{q})^a - \bar{p}^c (\bar{p}-\bar{q})_c \eta^{ab}}{\bar{p}^d \bar{p}_d (\bar{p}-\bar{q})^e(\bar{p}-\bar{q})_e},
\label{eq:pipiquase}
\end{align}
\end{widetext}
where $\bar{q}_a$ has the components
\begin{align}
\bar{q}_0&=q_0, \nonumber \\
\bar{q}_i &= \sqrt{g_{00}} N_i^{\;k}[ q_k + ({g}^{-1})_{kl} g^{0l} q_0],
\label{eq:q-bar}
\end{align}
with one similar expression for $\bar{p}^a$, in such way that
\begin{equation}
\bar{p}^d \bar{p}_d = p_0^2  + g_{00} \, \eta^{ij} \tilde{p}_i \tilde{p}_j.
\end{equation}
Performing the change of variables 
\begin{equation}
p_i \rightarrow \bar{p}_i,
\end{equation}
in the integral in Eq. \eqref{eq:pipiquase}, we obtain
\begin{widetext}
\begin{align}
\tilde{\Pi}^{ab}(\tilde{q}) &= e^2 2^{E(d/2)} g_{00} \sqrt{\frac{(-1)^{d-1} \bold{g}^{-1}}{ g_{00}^{d-1}}}T \sum_{n} \int \frac{d^{d-1}p}{(2\pi)^{d-1}} \frac{{p}^a({p}-\bar{q})^b + {p}^b ({p}-\bar{q})^a - {p}^c ({p}-\bar{q})_c \eta^{ab}}{\eta^{de}{p}_d {p}_e \eta^{fg}({p}-\bar{q})_f({p}-\bar{q})_g}\nonumber \\
&= \frac{\sqrt{|g|}}{g_{00}^{d/2-1}} \Pi^{ab} (\bar{q}),
\label{eq:pitotalfim}
\end{align}
\end{widetext}
where  $\Pi^{a b}(\bar{q})$ is the Minkowski space self-energy, as a function of the external momentum $\bar{q}$ \eqref{eq:q-bar}.

The static limit can be obtained assuming that 
all the components of the external momentum are negligible in the high temperature limit \cite{Frenkel:2009pi}. 
Then, Eq. \eqref{eq:pitotalfim} implies that the same is valid for $\tilde{\Pi}^{a b}$, so that
\begin{equation}
\left.\tilde{\Pi}^{ab}_{HTL} (\tilde{p}) \right|_{p_0 =0} =  \tilde{\Pi}^{ab} (\tilde{p}=0).
\label{eq:idparatildep}
\end{equation}
Using the result for the static self-energy in $d$ space-time dimensions \cite{Brandt:2012mu},  Eq.\eqref{eq:pitotalfim} yields
\begin{equation}
\left.\tilde{\Pi}^{ab}_{HTL} (\tilde{p}) \right|_{p_0 =0}  = -2 m^2 \tilde{u}^a \tilde{u}^b,
\label{eq:static-self-energy}
\end{equation}
where $m^2$ is the square of photon thermal mass, given by
\begin{align}\label{Tmass}
m^2 &=\sqrt{|g|} \left(\frac{T}{\sqrt{g_{00}}}\right)^2 \left[e^2 \left(\frac{T}{\sqrt{g_{00}}}\right)^{d-4} \right] \nonumber \\ &\times  \frac{2^{E(d/2)} (1-2^{3-d}) \Gamma(d-1) \zeta(d-2)}{\Gamma(\frac{d-1}{2})(2 \sqrt{\pi})^{d-1}} .
\end{align}
This thermal mass also behaves like a density under coordinate transformations (due to the factor $\sqrt{|g|}$) and depends on the temperature trough
$T_{\rm loc}$ defined by Eq. \eqref{Tloc}.

%

\newpage

\end{document}